# Overlap-Minimization Scheduling Strategy for Data Transmission in VANET

Yong Zhang[1,2,3]
1. *Institute of Advanced Technology on Communication, Chengdu Technological University*
Chengdu, China
2. *National Key Laboratory of Science and Technology on Communication, University of Electronic Science and Technology of China*
Chengdu, China
zhang.yong@uestc.edu.cn

Mao Ye[3]
3. *Department of Computer Science Loughborough University*
Loughborough, UK
M.Ye@lboro.ac.uk

Lin Guan[3]
3. *Department of Computer Science Loughborough University*
Loughborough, UK
L.Guan@lboro.ac.uk

*Abstract*— The vehicular ad-hoc network (VANET) based on dedicated short-range communication (DSRC) is a distributed communication system, in which all the nodes share the wireless channel with carrier sense multiple access/collision avoid (CSMA/CA) protocol. However, the competition and backoff mechanisms of CSMA/CA often bring additional delays and data packet collisions, which may hardly meet the QoS requirements in terms of delay and packets delivery ratio (PDR). Moreover, because of the distribution nature of security information in broadcast mode, the sender cannot know whether the receivers have received the information successfully. Similarly, this problem also exists in no-acknowledge (non-ACK) transmissions of VANET. Therefore, the probability of packet collisions should be considered in broadcast or non-ACK working modes. This paper presents a connection-level scheduling algorithm overlaid on CSMA/CA to schedule the start sending time of each transmission. By converting the object of reducing collision probability to minimizing the overlap of transmission durations of connections, the probability of backoff-activation can be greatly decreased. Then the delay and the probability of packet collisions can also be decreased. Numerical simulations have been conducted in our unified platform containing SUMO, Veins and Omnet++. The result shows that the proposed algorithm can effectively improve the PDR and reduce the packets collision in VANET.

*Keywords—VANET, CSMA/CA, overlap, scheduling, broadcast, overlay.*

## I. Introduction

Vehicular ad-hoc network (VANET) with inter-vehicle communication is an emerging technology for efficient safety and entertainment information dissemination. However, highly dynamic VANET topology due to the rapid mobility and varying vehicle density can cause unstable wireless links, and further lead to more packets loss and increased transmission delay, which bring challenges to guarantee the quality of service (QoS). The dedicated short-range communication (DSRC) [1] is a special wireless technology aiming at supporting both vehicle-to-infrastructure (V2I) and vehicle-to-vehicle (V2V) communications. DSRC also refers to a suite of standards including IEEE 802.11p, IEEE 1609.1/.2/.3/.4 protocol family and SAE J2735 message set dictionary[2]. DSRC is essentially a distributed communication system, and all connections in the system share the wireless channel, that is why a competition mechanism, carrier sense multiple access with collision avoidance (CSMA/CA), is adopted here. Unfortunately, the backoff mechanism of CSMA/CA will bring about the additional delay and possible packet collisions. Different from the performance improvement studies of VANET in [3] [4] [5] [6] [7] which mainly involve the upper-layer protocols, the study based on lower-layer protocol, such as the media access control (MAC) layer, is another research direction to improve the VANET performance. Compared with the optimization approaches implemented on the upper-layer protocols, the approaches on the lower-layer protocols have more efficient improvements.

In this paper, we propose a new approach about connections scheduling to improve the performance of data transmission in VANET, in which a connection consists of a sending node and a receiving node. The transmission means that a set of data packets are transmitted continuously from sender to receiver through a connection and transmission is also the scheduling unit of the proposed algorithm. The proposed approach overlays the existing MAC layer and CSMA/CA. Main contributions of this work are summarized as follows:

● Based on the VANET framework, a new overlay-based scheduling strategy is proposed to schedule the start-sending time of each transmission for maximizing the packet delivery ratio (PDR) and avoiding allocation issues of time slot.

● A transmission scheduling greedy search (TSGS) algorithm is proposed to avoid high time complexity.

## II. Relate Work

Due to the adoption of CSMA/CA in IEEE 802.11p standard, there are higher possibilities of packet collisions and unbounded delay in VANET, therefore there are a plethora of studies in the scope of MAC layer protocol to improve the performance of VANET, such as overlay. The overlay is an idea to perform resource configuration based on the existing MAC protocol. However, different from the IP layer which performs data encapsulation, the overlay only applies the existing MAC mechanism, e.g. CSMA/CA, to reconfigures resources, such as time, and not change MAC layer.

Herein, it is one study direction to combine the time-division multiple access (TDMA) and CSMA/CA for the

MAC-layer configuration in VANET. [8] proposed a demand-adaptive MAC (DAM) protocol based on dynamic TDMA and CSMA/CA. Once vehicles enter into the RSU's management domain (DM), they should shift to the listen status until getting the control message from the master. After broadcasting the control message, the vehicle could apply for a limited amount of time slots according to its demand. After the vehicle gets the authorization, it would send messages in its following time slots. Time is divided into periods called "round", and the duration time of each round could be diverse from each other. Each round has four stages: tell stage, ctrl stage, apply stage and information stage. On the basis of changeable network traffic, RSU can expand or shrink the duration of time slots dynamically round by round. [9] presented an application-oriented TDMA overlay MAC protocol by introducing an existing RA-TDMA[10] into the vehicular context to manage state sharing in the scope of collaborative applications. Each group of vehicles engaged in one collaborative application, e.g., a platoon or smart intersection, can form a team. RA-TDMA is then used to manage the periodic transmissions of each vehicle in the team for state sharing, and allocate them to different slots in a TDMA round with a fixed application-dependent period.

[11] considered that the existing approaches with bounded access delays cannot ensure a reliable broadcast service for safety applications and proposed an approach called QoS-aware centralized hybrid MAC (QCH-MAC). The proposal tries to combine TDMA and enhanced distributed channel access (EDCA) mechanisms together into a single architecture to improve their capabilities and to provide a higher QoS level. QCH-MAC is based on the idea that the access time can be divided into two periods: a transmission period (TP) and a reservation period (RP). TP consists of a set of TDMA slots called Tslot, while the TP period uses the EDCA protocol with two classes. The RP period is only employed by new vehicles to reserve their time slots in Tslot set. Different from IEEE 802.11p standard, the approach only considered two traffic classes: class1 and class2.

[12] proposed and verified empirical models of the relationship between network metrics and wireless conditions. Sufficient simulations were conducted with varying platoon sizes, numbers of occupied lanes and transmit power, to deduce the empirical models. Three different MAC protocols were involved in simulations, including the native CSMA/CA MAC of IEEE 802.11p standard and two overlay TDMA protocols (PLEXE-slotted and RA-TDMAp[13]) that can be readily implemented on IEEE 802.11p standard. In [14], the proposed communication protocol for platooning is also based on the IEEE 802.11p standard and TDMA. The rationale behind this protocol is to reduce random channel contention by adding the synchronization mechanism among nodes. Then, authors exploited the vehicle's position within the platoon to decide when to send messages for nodes. The idea in this paper is to divide the time into many slots after one beacon sent from the platoon leader, and to have each vehicle send its beacon in the time slot corresponding to its position in the platoon. This is different from a standard TDMA approach, as with TDMA node participating in the communication follow the same rules. However, in the proposed protocol, only nodes within a platoon cooperate in a TDMA-fashion.

In [13], the authors focused on the specific case of vehicles platooning applications and investigated the use of the RA-TDMA framework[10] on top of IEEE 802.11p standard to combine the benefits of both TDMA and CSMA/CA paradigms, namely collisions reduction through synchronization of beacons and efficient bandwidth usage with asynchronous access. The authors took inspiration from the technique used in [14] in which the leader of each platoon only transmits its beacons at high power, while the other platoon members use lower-power beacons and forward the information in a multi-hop scheme.

Except combining the TDMA and CSMA/CA to constitute overlay protocols, another study direction is to straight improve CSMA/CA mechanism. In IEEE 802.11p standard, the access category (AC) queues are applied to dispatch the priority according to various applications. Packets wait for an arbitrary inter-frame space (AIFS) before proceeding to the backoff stage. At the backoff stage, packets wait a random number of timeslots depending on both the contention window (CW) size of the backoff phase and the availability of the medium, before it is transmitted to the medium. Considering that CSMA/CA employs the conventional binary exponential backoff (BEB) scheme without consideration of medium's status, [15] originally proposed an adaptive backoff algorithm for the EDCA operation in IEEE 802.11p standard. The proposed algorithm takes into account the current status of the communication medium, evaluates the congestion level in the medium and uses this information to estimate the size of the backoff stage in the next transmission attempt. Owing to the fact that AIFS values in IEEE 802.11p standard are fixed and deterministic, [16] considered that the fixed values of AIFSs do not guarantee that ACs with higher priority can transmit over the medium before the lower one, and then proposed two algorithms to select AIFS values. The first algorithm guarantees strict priority to higher-priority ACs despite the value of CW. It eliminates the probability of a lower priority AC pre-empting a higher priority AC due to changeable AIFS values. The second algorithm is an adaptive non-deterministic algorithm that adjusts AIFS values according to the current status of the medium.

In contrast to previous work, our research focuses on the improvement based on the lower-layer protocol by overlaying MAC layer and CSMA/CA mechanism, meanwhile, the allocation and synchronization of time slots are not involved in the new approach.

III. SYSTEM FRAMEWORK AND PROPOSED ALGORITHM

In this paper, the system framework is shown in Fig.1. Vehicles (or called nodes) and roadside unit (RSU) are the entities that can send and receive data through wireless channels.

Assuming that RSU can communicate with all nodes within its coverage through control channels and all users send data packets through DSRC links. Obviously, the system works as the following hybrid network:

- Each sending-node, which intends to send data to other nodes, firstly sends its transmission requirement to RSU through the uplink of a control channel. The requirement includes data rate, deadline time to complete transmission, etc.

- RSU calculates the total duration of successive

packets in the transmission for each connection. By combining the duration and transmission requirement of sending-nodes, the RSU applies the proposed algorithm to obtain transmission schedule results.

- Through the downlink of a control channel, RSU sends scheduled results back to sending-nodes.
- According to the received schedule information, sending nodes transmit their data packets to their destination through DSRC links.

As the aforementioned procedure, it is much important to develop a scheduling algorithm that can produce a good schedule to meet the nodes' requirements about transmission. In this paper, we have proposed an algorithm to fulfil the mentioned issues, and it is quite easy to integrate the algorithm into RSU.

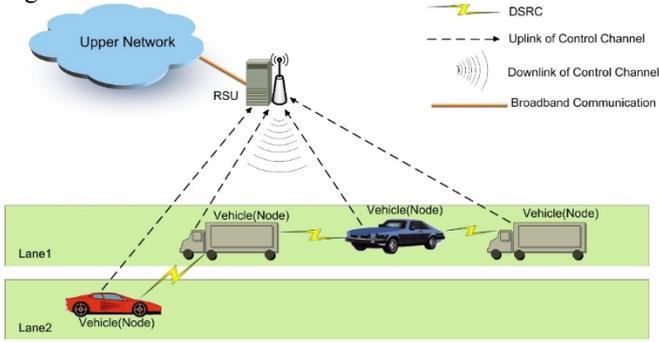

Fig.1 System Framework

In VANET, the packet delivery ratio (PDR) is an important metric of performance. In this work, we aim to maximize the PDR of VANET with the least amount of collision. The objective function can be expressed as

$$\Omega = \arg\min_{t_i \in T_i} f(\boldsymbol{t}) \quad (1)$$

Subject to $T_i + \tau \leq Q_i, i = 1, 2, ..., N$

where $T_i$ is the available sending-time window. $Q_i$ is the sending deadline which meets the transmission requirement of the $i^{th}$ connection. Herein, a connection means a direct peer-to-peer link between two nodes. $N$ is the number of all connections to be scheduled. $\tau$ is an additional margin for possible delay caused by CSMA/CA backoff, without loss of generality, $\tau = 0$ in this paper. $\boldsymbol{t} = \{t_1, t_2, ..., t_N\}$ is a vector consisting of scheduled sending time for each connection, as shown in Fig.2 (a). Each connection involves a sending node and a receiving node. Duration is the total period during which the sender continuously sends its data packets over the wireless channel.

$f(\boldsymbol{t})$ in formula (1) is a cost function measuring the overlap between durations, and it is defined by

$$f(\boldsymbol{t}) = \sum_{i=1}^{N} \sum_{j=1(j \neq i)}^{N} \varepsilon_{ji} \quad (2)$$

The definition of cost function $f(\boldsymbol{t})$ is based on the fact that less overlap of durations can reduce collision and time delay. That is because each duration in Fig.2 consists of multiple packet periods and less overlap of durations means that fewer vehicles stay in the backoff stage at the same time. $\varepsilon_{ji}$ is the time overlap between duration $d_j$ and duration $d_i$ as shown in Fig.2 (b). Therefore, the object to maximize PDR can be converted to minimize the overlap between durations.

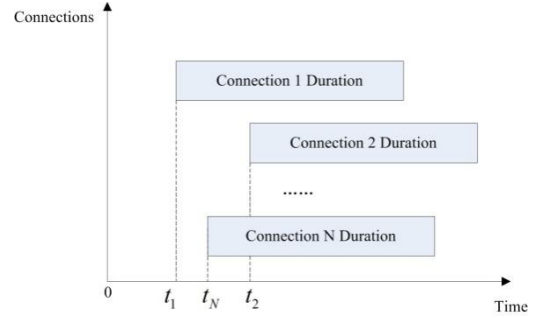

(a)

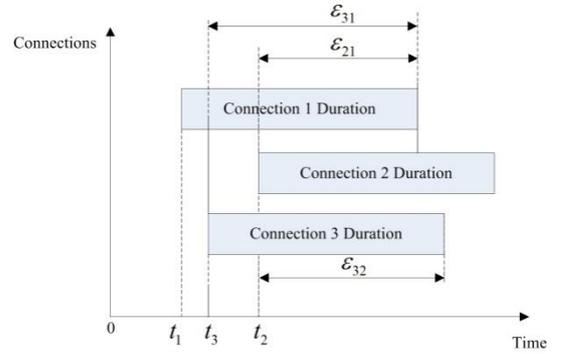

(b)
Fig. 2 Variables Demonstration

The computational complexity of formula (2) is easily calculated as $o(N^2)$. To get the optimum solution of formula (1), the search time step $\delta$ for $t_i$ must be introduced. According to the size of $\delta$, duration $[0 - t_i]$ is divided into $M = t_i / \delta$ segments. Given the number of searches as $M$, it means each connection has $M$ possible schedules. When $N$ connections are considered in the system, the time complexity of formula (1) is $o(M^N)$, which is very high. To alleviate the complexity of formula (1), in this paper, we have proposed an algorithm called transmission schedule greedy search (TSGS) as follows.

TABLE I. MAIN NOTATIONS USED IN TSGS ALGORITHM

| Notations | Definitions |
|---|---|
| $N$ | The number of all connections in the system |
| $\boldsymbol{Q}$ | According to the transmission requirements of each connection, define $\boldsymbol{Q}$ as a set of deadline time |
| | $\{q_1, q_2, ..., q_N\}$, whose element indicates the deadline time when packets transmissions of each connection have been completed. |
| $W$ | Set of time windows $\{w_1, w_2, ..., w_N\}$ whose element is the available time period for scheduling. |
| $t$ | Set of scheduled sending time $\{t_1, t_2, ..., t_N\}$, |
| $D$ | Set of connection durations $\{d_1, d_2, ..., d_N\}$ |
| $\varepsilon_{i,j}$ | The size of time overlap between $d_i$ and $d_j$ |
| $\delta$ | Search step of time when scheduling |

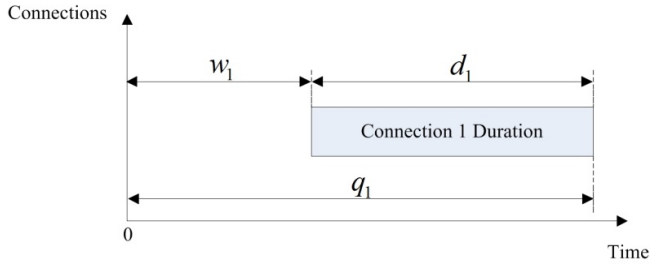

Fig. 3 Time-related variables demonstration

---

**Algorithm:** TSGS

**Input:** the transmission requirements $Q$

1. **While ( $i \leq N$ )**

    // Get the available time windows $W$ for scheduling, as shown in Fig.3.

2. $w_i = q_i - d_i$

3. **While ( $i \leq N$ )**

    // $\sigma$ is the number of searching in step $\delta$ for the $i^{th}$ connection

4. $\sigma = w_i / \delta$

5. $k = 0$

6. **While ( $k < \sigma$ )**

    // $J$ is the number of scheduled connections before connection $i^{th}$, so $J = i - 1$

7. **While ( $j \leq J$ )**

    // Get the cost measurement, i.e. the overlap.

8. $\Phi_i(k) = \sum \varepsilon_{i,j}(k)$

9. $k = k + 1$

    // Get the search index with the minimal scheduling cost

10. $k_{min} = \underset{k \in [1,...,\sigma]}{\arg\min}(\Phi_i)$

    // Convert the index to sending time

11. $t_i = k_{min} \times \delta$

**Output:** $t = [t_1, t_2, ..., t_N]$ // Finished scheduling and obtain sending time for each connection.

---

The proposed TSGS algorithm schedules each upcoming connection $i$ based on the scheduled $J$ connections as the description in the pseudo-code program. Each new connection only considers its $M$ possible schedules and the scheduled connections remain unchanged, therefore the time complexity of TSGS algorithm is $o(MN)$ which is much lower than that of formula (1).

## IV. PERFORMANCE EVALUATION

To investigate the performance of the proposed TSGS algorithm, we conduct a set of simulations on a unified platform which consists of a road network editor for OpenStreetMap, a road traffic generator SUMO, a network simulator Omnet++, and a vehicular network framework Veins.

### A. Simulation Setup

By using OpenStreetMap (OSM), we extract the road map of M1 highway near Loughborough University, UK, in Fig. 4. On the left-hand side, it is the original map, whereas the right-hand side is the extracted road map from OSM. In this work, we assume all vehicles in the scenario can error-freely communicate with the scheduling controller deployed in RSU.

We use SUMO to generate a traffic flow, which includes trajectories of 7 vehicles. 4 of them are selected to generate 2 connections. SUMO generates the trajectories of vehicles by following some rules, such as speed limitation (45 miles/h). The simulation time is 20 seconds.

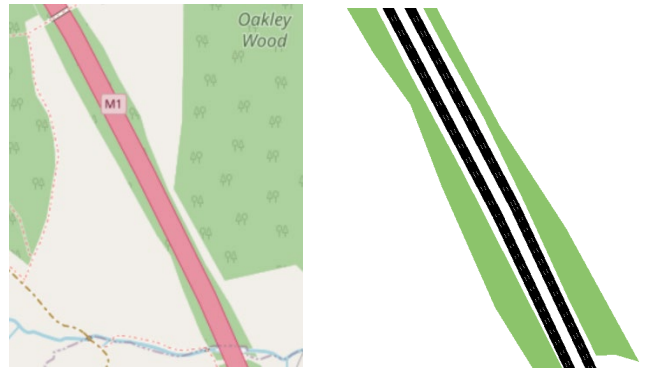

Fig.4 M1 highway map near Loughborough University, UK

The setting of important parameters for the numerical experiment has been as shown in Table II.

TABLE II. THE SETTING OF IMPORTANT PARAMETERS

| Parameter | Value |
| --- | --- |
| Transmission Power | 0.5W |
| Speed limit | 45 mph |
| Date rate of DSRC | 6.0 Mbps |
| Bandwidth of DSRC | 10.0 MHz |
| Noise floor | -98dBm |
| Wireless channel model | Simple path loss |
| Frequency of wireless channel | 5.9 GHz |
| DSRC Transceiver for each vehicle | 1 |
| Control signalling Transreceiver for each vehicle | 1 |
| Packets for each connection | 50 |
| Time duration of a packet | 23us |
| Scheduling controller (RSU) | 1 |
| connections | 2 |
| vehicles (nodes) | 7 |

### B. Simulation Results

This set of experiments compares the performance of the proposed TSGS scheduling algorithm with random sending. Random sending is in line with the actual situation in most distributed systems. Since we work for maximizing the PDR and the probability of successfully received packets is the performance objective of this work, PDR and received packets are respectively selected to make comparisons.

Fig.5 shows the overall PDR of the scenario and the number of received packets for each connection with the gradually increasing size of available schedule time windows. Each connection is a random wireless link between two nodes. The available schedule time window of a connection is determined by its transmission time requirement, in fact, it is the time window allowed to transmit packets as the $q_1$ shown in Fig.3. Without loss of generality, we assume that the transmission time requirements of all connections are the same in our work, thus schedule time windows have the same size. As shown in Fig.5, the proposed TSGS schedule algorithm gives better performance than the random sending mechanism on both PDR and the number of received packets. Even in the two-connections scenario, the total average PDR of the proposed algorithm is 86.11%, which is higher about 10% than 76.33% the average PDR of random sending. Also, the proposed algorithm has similar improvements to the number of received packets. It can be easily imagined that our proposed algorithm should be much superior to the random sending mechanism in the scenarios with more connections or more complex systems.

As the schedule time window increases, the PDR and the number of received packets almost increase monotonically for the proposed algorithm. However, random sending has not such a performance. That is because as the size of schedule time window increases, smaller or even no overlap among durations can be achievable by applying the TSGS schedule algorithm, as shown in Fig. 6 (a). As shown in Fig.6 (b), much duration overlap activates the backoff mechanism of CSMA/CA and causes greater transmission delay for each connection. That is because the time gap between packets in each connection in Fig.6 (b) is bigger than that in Fig.6 (a). Here, the abscissa (horizontal coordinate) is time, but the ordinate (vertical coordinate) has no specific meaning, just in order to distinguish the two connections. It is worth mentioning that Fig.6 is independent, just to show the idea of TSGS algorithm, and has nothing to do with other graphs.

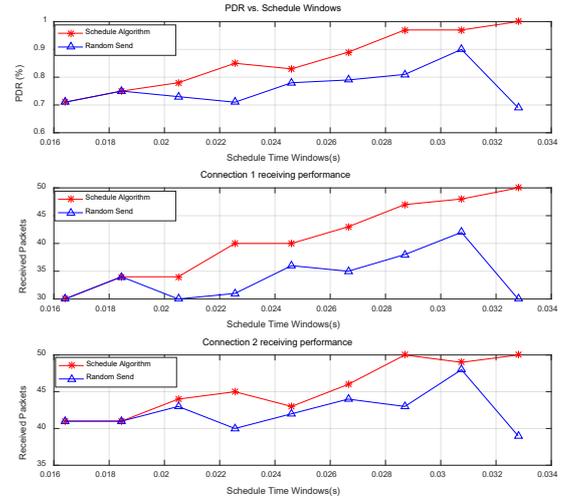

Fig.5 The PDR and received packets with an increased schedule time windows

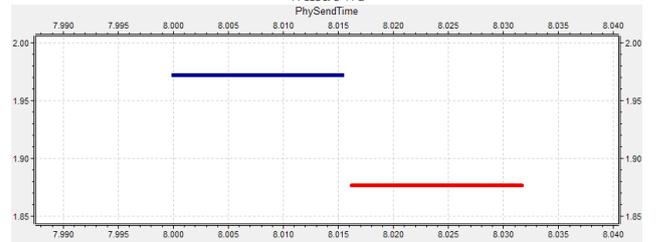

(a) Durations of TSGS schedule algorithm

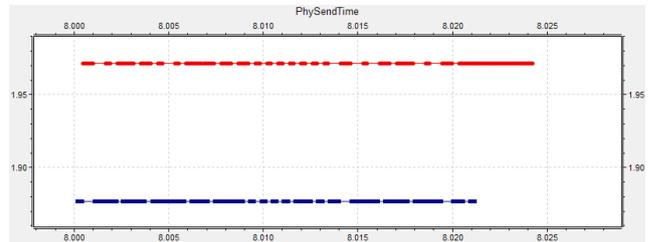

(b) Durations of random sending

Fig.6 Durations demonstration of 2 connections
(Transmission requirement: Time window to complete transmission is 8.0s~8.0328s. 50 packets are sent continuously in each connection.)

A vehicle, receiving data packets through a connection, cannot successfully process collided packets in the broadcasting situation due to the lack of acknowledgement. As shown in Fig.7, the collision feature of the proposed algorithm is compared with that of random sending. The proposed algorithm can achieve better performance on collision avoidance. When the schedule time window reaches 0.0328s, there has been no packet collision by using TSGS schedule algorithm. However, random sending will still cause quite a bit of collision. Additionally, one collision involves at least two packets which belong to two different connections, therefore, two related connections, connection 1 and connection 2, have the same collision performance as shown in the last 2 sub-graphs of Fig.7.

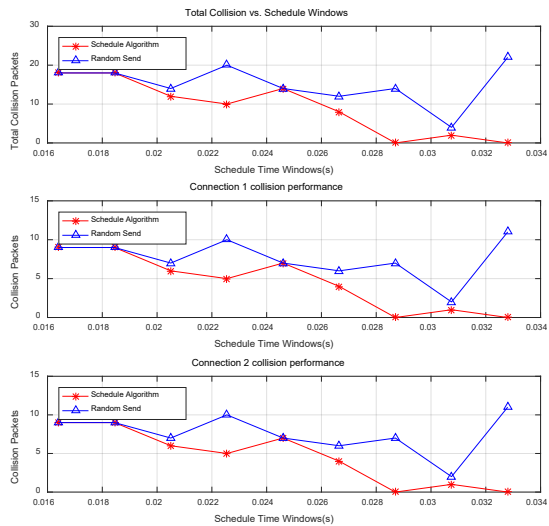

Fig.7 Collision performance

## V. CONCLUSION

In this paper, we are committed to solving the problems in the distributed VANET. Due to the lack of global information, the packets collision concerned by researchers will frequently arise in VANET scenarios. It can be summarized by our work that if there are fewer vehicles in the backoff phase at the same time, the probability of collision will be lower. It can be further concluded that reducing the overlap between connections can effectively alleviate the collision when each connection consists of multiple consecutive packets transmissions. Based on the VANET framework, a connection-level scheduling strategy is proposed to schedule the packets transmissions for all connections by minimizing their overlaps. Then, a related algorithm named TSGS is presented to find the optimal sending time for each connection with higher PDR, meanwhile meeting the transmission requirement of each connection. Extensive simulation experiments have been conducted to compare the proposed algorithm with random sending. The experiment results show that the average PDR of the proposed algorithm is higher about 10% than that of random sending. Moreover, the proposed algorithm is easily embedded in RSU and combined with VANET routing schemes. In the future, the software-defined network (SDN) technology can be involved in RSU to carry out central control with global information.

ACKNOWLEDGEMENT

Authors thank the funding of Science and Technology Program of Sichuan Province (2018JY0507)& (2020YFH0071), Natural Science Foundation of China Project (61871422).